\documentclass[twocolumn,showpacs,aps,floatfix]{revtex4}

\usepackage{amsmath}
\usepackage{graphicx}
\usepackage{dcolumn}
\usepackage{bm}

\begin{document}

\title{Parametrical CI+MBPT calculations of Th I energies and $g$-factors for even states}
\author{I. M. Savukov}

\affiliation{Los Alamos National Laboratory, Los Alamos, NM 87544,
USA}

\date{\today}

\begin{abstract}
In this paper, we study energies and $g$-factors for a large number of levels of Th I. We found that the accuracy of CI-MBPT can be substantially improved by using multiple adjustment parameters, which are introduced to regulate the second-order MBPT corrections for single valence energies and Coulomb interaction. The results for energies and $g$-factors are in excellent agreement with experiment, with accuracy sufficient to positively assign theoretical states of given angular momentum and parity to experimental levels. This theory will be further developed in the future to treat other, even more complicated actinides such as U I.

\end{abstract}
 \maketitle
\section{Introduction}

Many applications require data on energy levels, transition probabilities, lifetimes, and oscillator strengths of complex actinide atoms, which  are currently lacking. Theory can be used to supply missing data; however, the existing theories are unable to saturate large valence configuration space and include strong valence-core interactions and relativistic effects, so their accuracy is limited in most actinide atoms and low-charge ions. For developing theory it is very important to test it in a well characterized actinide atom, such as Th I, which has a large number of energy levels available with assigned J, parity, and dominant configurations and with experimentally measured $g$-factors.  With only four valence electrons, Th I is also an atom of moderate complexity for which calculations can be performed in a reasonable time.  Apart from the needs of testing theory, the Th I spectrum is of certain interest owing to the possibility of development of a nuclear clock \cite{Peik} and its conventional use as the standard for high-resolution spectroscopy \cite{Redman}.

Previously, an {\it ab initio} configuration-interaction many-body perturbation theory (CI+MBPT) method, which can achieve high precision in multi-valence atoms, was applied to calculations of energy levels for many low-lying levels of Si I, which has four valence electrons as Th I, but with simpler core, and good agreement was observed \cite{SiCIMBPTen}. In addition to Si I \cite{SiCIMBPTen}, energy calculations with the CI+MBPT method were performed for other 4v atoms such as C \cite{Berengut} and Ge, Sn, Pb \cite{DzubaGe}. However according to our preliminary calculations, the {\it ab initio} CI-MBPT approach appears to have much lower accuracy in Th I, insufficient for obtaining energy levels in correct order which creates difficulties in identification and further applications of the theory. The main reason for this is limited treatment of strong valence-core interactions only in the second order and relativistic effects particularly strong in heavy atoms. When valence core interaction is treated in all orders and relativistic effects are included more consistently, as in configuration-interaction (CI)-all order calculations \cite{ThSafronova}, the accuracy is significantly improved.  Although for Th I the CI-all-order approach is quite appropriate, it requires large computational resources and is quite slow, while its accuracy needs yet to be demonstrated for more complex actinides. Also the CI-all-order calculations in Th I were limited to a small subset of low-lying energy levels.

 However, as we will show here, it is possible to significantly improve CI-MBPT theory accuracy by using nine adjustable parameters: accuracy about 200 cm$^{-1}$ can be achieved when 16 levels of the same total angular momentum J are fitted. This accuracy is sufficient for straightforward identification for many lowest levels. In addition, it is possible to use the same fitting parameters for a given subset of levels of specific J and parity to reproduce much larger number of levels, without additional fitting. This way the theory has predictive power for energy levels and can be used to check if there are some missing experimental levels. In this work we focused on fitting Th I 16 levels for each J (even states) for the purpose of demonstrating the method. The theory can be applied to other atomic and ionic systems. Land\'e $g$-factors, after the parameters are optimized by fitting the energy levels, are also quite accurately predicted, in fact  with similar accuracy as the CI-all-order method \cite{ThSafronova}. This is significant since the $g$-factors when accurately calculated help to remove  ambiguity in the identification and indicate correctness of mixing of configurations of different types.
 
 Parametric fit methods have a long history of applications. It started with a work by Giulio Racah in Physica (16, 1950, 651) on a parametric fit
for spectrum of Th III.

 While the optimization might appear as a simple standard numerical task, it is not in the case of nine fitting parameters used in CI-MBPT. First of all, the CI-MBPT calculations even for Th I take quite a long time and in order to find a minimum it is necessary to repeat calculations 100 times with different parameters. It is still an open question what is the best algorithm and whether the absolute minimum is reached, but here we at least found one practical algorithm described below  that allows to obtain about 200 cm$^{-1}$ accuracy.

Once the parameters have been optimized, various properties, such as $g$-factors, can be calculated. The question can be asked whether the accuracy of wavefunctions of a large-scale {\it ab initio} theory is improved by introducing multiple fitting parameters. Usually {\it ab initio} approaches are highly desirable, because their high accuracy for energies  leads to high accuracy for other properties, such as for example oscillator strengths; however,
 owing to poor convergence of MBPT and requirement of very high accuracy of level splittings to have correct mixing coefficients, the utility of pure {\it ab initio} theory can be limited.  In particular, higher energy levels are extremely sensitive to small corrections and it is  impractical to achieve sufficient accuracy by using very large basis sets and high-order MBPT corrections. One specific observable is the $g$-factor, which is sensitive to the mixing coefficients of different types of configurations. Thus by testing theoretical predictions for $g$-factors, it is possible to evaluate the performance of the theory when fitting is used for energies only. Of course it is further possible to improve the accuracy of $g$-factors as well by including them into the fitting procedure; however, then they cannot be used to test the theory. In this work we focus on fitting 9 parameters to minimize the deviations of energies for 16 levels and then we test the energies and $g$-factors.

\section{Method}
To calculate Th I energies a CI+MBPT method developed for open shell atoms with multiple valence electrons is used (see for example \cite{DzubaGe}). The theory can be summarized as follows. The effective CI+MBPT Hamiltonian for Th I is split into two parts:
\begin{equation}
H^{eff}=\sum_{i=1}^{M} h_{1i}+\sum_{i\ne j}^{M} h_{2ij}.
\end{equation}
The one-electron contribution
\begin{equation}
 h_{1}=c\mathbf{\alpha}\cdot\mathbf{p}+(\beta-1)mc^2-Ze^2/r+V^{N-4}+\Sigma_1
 \end{equation}
in addition to the $V^{N-4}$ Dirac-Hartree-Fock (DHF) potential contains the valence electron self-energy correction, $\Sigma_1$ \cite{Dzuba1987}.  In the current CI+MBPT program, the self-energy correction is calculated with the second-order MBPT. The term $\Sigma_1$ is regulated with seven scaling factors each for a specific one-electron relativistic angular momentum number: $s_{1/2}$, $p_{1/2}$, $p_{3/2}$, $d_{3/2}$, $d_{5/2}$,$f_{5/2}$, $f_{7/2}$. These factors not only take into account some omitted high-order MBPT corrections, but also relativistic effects such as single-particle Breit terms.
The two-electron Hamiltonian is
\begin{equation}
h_2=e^2/|\mathbf{r_1}-\mathbf{r_2}|+\Sigma_2
\end{equation}
where $\Sigma_2$ is the term accounting for Coulomb interaction screening arising from the presence of the core \cite{Dzuba89}. In the CI-MBPT program used, the screening is also calculated in the second order. For fitting the two additional scaling factors are introduced for zero and first-order multipolarity of the Coulomb interaction. Further details on the CI+MBPT approach can be found in Ref.\cite{Dzuba96}.
In terms of specific numerical steps, first, the DHF V$^{N-4}$ potential for the closed-shell Th V ion is calculated. Second, the basis in the frozen V$^{N-4}$ potential is calculated with the help of a B-spline subroutine for the ion in a cavity of radius $R$. The basis is then used to evaluate the CI+MBPT terms in Eq. 1. Finally, the eigenvalue problem is solved for the effective Hamiltonian matrix.
The program can generate a set of configurations by single-~, double-, etc. excitations of the input configurations limited by a given maximum angular momentum $l_{max}$ and $N_{max}$. In case of Th I, the number of states grows very fast, so the reference states are carefully chosen from which only very limited number of excitations is allowed. For uniform treatment of all J states, the same reference states and excitation protocol are used. The effective Hamiltonian matrix generation is repeated multiple times for different scaling factors (9 total) and optimization procedure described below is used until some level of convergence is reached.


\section{CI+MBPT Th I energy calculations}
In order to rigorously test the accuracy of CI+MBPT method and the convergence patterns for the fitting parameters, we made extensive calculations of a large number of energy levels. Sixteen states were chosen for the fitting procedure for nine parameters for J ranging from 1 to 6. We excluded fitting for $J=7$ because 16 levels for J=7 would lead to very high energies. Instead we used parameters obtained by fitting J=6 levels to obtain J=7 levels and compared the resulting energies with experiment. We found excellent agreement, which is an indication of predictive power of the method for energy levels. In addition, in case of J=6 we used the theory with nine fitted parameters for  16 J=6 levels to produce higher levels not included into the fitting and found that many experimental levels seem to be missing.
\subsection{Choice of configurations}
Because using a large basis set that can saturate the valence-valence CI is of high computational cost for Th I, we tried to find a most efficient small subset of configurations that can account for most important CI effects. In case of $V^{N-4}$ basis calculations adopted in this paper, the radial wavefunctions of single-electron states, unlike MCHF calculations where radial wavefunctions can be optimized, deviate significantly from the physical one, so it is necessary to include configurations that would correct radial functions of zero order. To include a large number of radial states is quite impractical for 4v systems, so instead, the configurations were limited by inclusion only of the next state in the radial quantum number $n$: 7s, 8s, 5f, 6f, 6d, 7d. By inspecting the configuration weights, we observed that higher n states have very small weights and hence, at the accuracy aimed here, can be neglected especially since their effect can be partially taken into account by adjusting fitting parameters. Thus for Th I, we started with basic configurations such as $6d^27s^2$, $6d^37s$, $5f7s^27p$, $6d^4$, $5f6d7s7p$, and included lowest single excitations: $6d^27s8s$, $6d7d7s^2$, $6d^38s$, $6d^27d8s$, $5f7s^28p$, $5f7s8s7p$, $6f7s^27p$, $6d^37d$, $5f6d7s8p$,$5f6d8s7p$,$5f7d7s7p$,$6f6d7s7p$.  Single and double excitations from these listed non-relativistic configurations were also allowed to 8s, 8p, 7d, 6f.
\subsection{Fitting parameters}

The initial values of fitting parameters were found for $k_1$,$k_2$,...,$k_7$ [$\Sigma_1(s_{1/2})$,$\Sigma_1(p_{1/2})$,...,$\Sigma_1(f_{7/2})$] by minimizing deviations between experimental and theoretical Th IV energies (Table \ref{TableIa}). In case of $s$  as well as $p$ states, the optimization for the $n=7$ states (difference between the ground 5$f_{5/2}$ and 7$s$ or 7$p$ energies) did not result in small deviation for the $n=8$ states. Thus although the single-valence energies calculated in $V^{N-4}$  potential might be the same in Th I, II, III and IV, there are different mixtures of $n=7$ and $n=8$ single-valence orbitals, when those are included in CI expansion, resulting in variations of the optimal $k_1$,$k_2$,...,$k_7$ parameters, so further optimization is needed for each specific ion, J, parity, and the range of energy levels. Similar variation is expected  for the parameters $k_8$ and $k_9$, respectively the monopole and dipole Coulomb screening terms $\Sigma_2(L=0)$ and $\Sigma_2(L=1)$. The initial approximation for these parameters was found by optimizing theoretical energies for  Th II with the other seven parameters fixed. The resulting set of parameters:  0.6330, 0.6330, 0.5820, 0.7860, 0.8010, 0.9170, 0.9250, 1.2188, 0.6978 was used as an initial guess for minimization of deviations with experiment for J=6 even states of Th I with the following algorithm:

1) a minimum was determined for the first parameter;

2) its value was substituted and the minimum was determined for the second parameter;

3) the procedure was repeated until all nine parameters were optimized in sequence;

4) their values were substituted and the steps 1-3 were repeated until the changes in the energy deviations became small.

After the optimization had been completed for J=6 states, the optimal parameters were used as an initial guess for the $J=5$ optimization. The procedure was repeated in sequence for $J=4,3,2,1$.
The optimization steps 1-4 required many iterations. The sets of optimal parameters for $J$-specific level systems are shown in Table~\ref{TableII}. The residual deviation in the fit is listed in the last raw of the table, which for $J=6$ case is quite small, 159 $cm^{-1}$. One important observation is that the parameters significantly deviate from 1, meaning large corrections to second-order MBPT. The other interesting feature is that the parameters significantly vary for different J. Still the variation in each parameter value is  smaller than deviation from unity in most cases. Although it is possible to run optimization without limitations on the value of parameters, it is important to restrict them by some values to avoid unrealistic MBPT corrections. Specifically, we restricted parameters to positive values and tried to keep them close to the initial guess. However, in some cases, the changes were quite large due to relatively small effect of the specific parameters on the energies. In order to understand how the accuracy depends on the deviation of parameters from expected realistic values, the results for two much different sets that led to similar deviations of theoretical energies and g-values from experiment and to similar leading configuration percentages are compared (Table \ref{TableIII}). This appears to indicate that even unrealistic fit parameters can still lead to reasonable wavefunctions. However, in case of more physical parameters some smooth behaviour between different J states can be observed so they can be used to calculate energies for different J than used in the fitting.

\begin{table}
\caption{Optimization of parameters $k_1$,$k_2$,...,$k_7$ [$\Sigma_1(s_{1/2})$,$\Sigma_1(p_{1/2})$,...,$\Sigma_1(f_{7/2})$] by matching the energies for  Th IV.
Because theory for the lowest and the next states for a given $l$ and $j$  does not match experiment with one value of the corresponding parameter, two values are given in case of s and p states which have larger deviations between theoretical and experimental energies, $\Delta{E}$. From this fact it follows that the used fitting parameters in multi-valence atoms cannot be the same for different J values owing to differences in configuration expansions, that is content of e.g. 7s and 8s states \label{TableIa}}
\begin{ruledtabular}
\begin{tabular}{crrrlc}
States	&	$E_{exp}$	&	$E_{th}$&$\Delta{E}$	&	$k_i$	&	$i$	\\
\hline
$5f_{5/2}$	&	0	&	0	&	0	&		&		\\
$6f_{5/2}$	&	127269	&	127273	&	-4	&	0.917	&	6	\\
$5f_{7/2}$	&	4325	&	4322	&	3	&	0.925	&	7	\\
$6f_{7/2}$	&	127815	&	127718	&	97	&	0.925	&	7	\\
$6d_{3/2}$	&	9193	&	9195	&	-2	&	0.786	&	4	\\
$7d_{3/2}$	&	119685	&	119569	&	116	&	0.786	&	4	\\
$6d_{5/2}$	&	14486	&	14486	&	0	&	0.801	&	5	\\
$7d_{5/2}$	&	121427	&	121260	&	167	&	0.801	&	5	\\
$7p_{1/2}$	&	60239	&	60237	&	2	&	0.633	&	2	\\
$8p_{1/2}$	&	134517	&	134095	&	421	&	0.633	&	2	\\
$7p_{1/2}$	&	60239	&	61594	&-1355	&	0.460	&	2	\\
$8p_{1/2}$	&	134517	&	134505	&	11	&	0.460	&	2	\\
$7p_{3/2}$	&	73056	&	73055	&	1	&	0.582	&	3	\\
$8p_{3/2}$	&	139871	&	139414	&	457	&	0.582	&	3	\\
$7p_{3/2}$	&	73056	&	74434	&-1378	&	0.362	&	3	\\
$8p_{3/2}$	&	139871	&	139870	&	0	&	0.362	&	3	\\
$7s_{1/2}$	&	23131	&	23132	&	-1	&	0.632	&	1	\\
$8s_{1/2}$	&	119622	&	119178	&	443	&	0.632	&	1	\\
$7s_{1/2}$	&	23131	&	24814	&-1683	&	0.478	&	1	\\
$8s_{1/2}$	&	119622	&	119619	&	3	&	0.478	&	1	\\
\hline
\end{tabular}
\end{ruledtabular}
\end{table}

\begin{table}
\caption{Optimized parameters (with attempts to keep them close to realistic values obtained from Th IV) for different even J states and residual deviation ($\sigma$) for 16 states of specific J in cm$^{-1}$.  \label{TableII}}
\begin{ruledtabular}
\begin{tabular}{llcccccc}

 Par. &J=1 & J=2 & J=3 & J=4  & J=5 & J=6\\
 \hline
1 &0.5740 &0.5535 &0.6022 &0.5142 &0.5232 &0.6724 \\
2 &0.6307 &0.7101 &0.5813 &0.4557 &0.5257 &0.5258\\
3 &0.7435 &0.6328 &0.5940 &0.9318 &0.9958 &0.3172 \\
4 &0.7430 &0.7552 &0.7517 &0.8192 &0.8052 &0.8556 \\
5 &0.8301 &0.8058 &0.8175 &0.8461 &0.8817 &0.8693 \\
6 &0.8201 &0.8138 &0.8211 &0.8100 &0.8151 &0.8076 \\
7 &0.9860 &1.0215 &0.8967 &0.8518 &0.8466 &0.8780 \\
8 &1.1653 &1.1794 &1.2935 &1.2983 &1.2544 &1.3265 \\
9 & 0.0291 &0.0000 &0.0000 &0.0000 &0.0000 &0.5195 \\
\hline
$\sigma$& 275& 230  & 250 & 213   & 258   & 159 \\

\end{tabular}
\end{ruledtabular}
\end{table}

\begin{table}
\caption{Comparison of two leading configuration expansion coefficients and $g$-factors obtained for two very different sets of optimized parameters for J=6 even states with those in Ref.\cite{Thorium}.
Set 1 is given in Table \ref{TableII} and Set 2 is: 0.7089, 1.5880, 1.9052, 0.8371, 0.9422, 0.5566, 0.6278, 2.1353, 0.4924.
  \label{TableIII}}
\begin{ruledtabular}
\begin{tabular}{lrrrrrr}
St. 	&	Set 1	&	Set 2	&	Ref.\cite{Thorium}	&	$g_{Set 1}$	&	$g_{Set 2}$ 	&	$g_{expt}$	\\
	&	 Conf. $\%$	&	Conf. $\%$ 	&	Conf. $\%$ 	&		&		&		\\
\hline
1	&	83 sd$^3$	&	82 sd$^3$ 	&	98 sd$^3$	&	1.167	&	1.167	&	1.165	\\
    &   2 d$^4$      &   2 d$^4$     &  2 d$^4$      &           &           &           \\
\hline
2	&	90 spdf	    &	85 spdf	&	spdf	&	1.1174	&	1.1232	&	1.110	\\
    &   6 pd$^2$f    &   7 pd$^2$f  &          &           &           &           \\
\hline
3	&	76 d$^4$   	&	70 d$^4$	&	99 d$^4$	&	1.1294	&	1.1351	&	1.125	\\
    &   9 pd$^2$f      &   12 spdf &           &           &           &           \\
\hline
4	&	89 spdf	&	87 spdf	&	spdf	&	1.2053	&	1.2078	&	1.185	\\
    &   4 pd$^2$f      &   6 pd$^2$f &          &           &           &           \\
\hline
5	&	88 spdf	&	85 spdf	&		&	1.1418	&	1.1548	&	1.155	\\
    &   8 pd$^2$f      &  8 pd$^2$f  &          &           &           &           \\
\hline
6	&	87 spdf	& 83 spdf	 &	spdf	&	1.2428	&	1.252	&	1.200	\\
    &   4 d$^4$        &  4 pd$^2$f&          &           &           &           \\
\hline
7	&	48 spdf	& 67 d$^4$	&	100 d$^4$	&	1.1713	&	1.0476	&		\\
    &   35 s$^2$f$^2$        &  10 spdf&          &           &           &           \\
\hline
8	&	74 d$^4$	&	46 spdf	&		&	1.0581	&	1.1715	&	1.210	\\
    &  9 pd$^2$f        &  27 s$^2$f$^2$&          &           &           &           \\
\hline
9	&	67 spdf	&	59 spdf	&		&	1.0785	&	1.0695	&	1.080	\\
    &  27 pd$^2$f       & 35 pd$^2$f&          &           &           &           \\
\hline
10	&	73 spdf	&	73 spdf	&		&	1.1869	&	1.1822	&	1.205	\\
    & 11 pd$^2$f      & 11 pd$^2$f&          &           &           &           \\
\hline
11	&	82 spdf	&	81 spdf	&		&	1.2216	&	1.217	&	1.190	\\
    & 6 s$^2$f$^2$      & 5 s$^2$f$^2$&          &           &           &           \\
\end{tabular}
\end{ruledtabular}
\end{table}

\begin{table*}
\caption{Comparison of experimental energies $E_{ex}$ and g factors $g_{ex}$ given in \cite{Thorium} with theoretical energies $E_{th}$ and g factors $g_{th}$ for CI-MBPT theory with 9 adjustable parameters optimized for  first 16 even J=6 levels. Energies are given in cm$^{-1}$;  $\Delta E $ is the difference between experimental and theoretical energies, which are aligned for the first level given in the table; $\Delta g$ is the difference between theoretical  and experimental g factors.   \label{Table1}}
\begin{ruledtabular}
\begin{tabular}{rcrlclrccrcllclccc}
$\#$ &  $E_{ex}$  & Conf.\cite{Thorium}         &  $g_{ex}$      &   $E_{th}$& Conf.[th]         & $\Delta E $   &   $g_{th}$  &  $\Delta g$       &$\#$ &  $E_{ex}$  & Conf.\cite{Thorium}         &  $g_{ex}$      &   $E_{th}$& Conf.[th]         & $\Delta E$   &   $g_{th}$  &  $\Delta g$  \\
\hline
1	&	16554	&	$6d^37s$	&	1.165	&	16554	&	$6d^37s$	&	0	            &	1.167	&	-0.002	&	17	&	37742	 &		&	        	&	37778	&	$5f6d7s7p$	&	-36	&	1.018	&		\\
2	&	26997	&	$5f6d7s7p$	&	1.11	&	26855	&	$5f6d7s7p$	&	142	&	1.117	&	-0.007	&	18	&		&		&		&	38933	&	$5f6d7s7p$	&		&	1.092	&		\\
3	&	27972	&	$6d^4$	&	1.125	&	27908	&	$6d^4$	&	64	&	1.129	&	-0.004	&	19	&		&		&		&	39480	&	$5f6d^27p$	&		&	0.988	&		\\
4	&	29553	&	$5f6d7s7p$	&	1.185	&	29603	&	$5f6d7s7p$	&	-50	&	1.205	&	-0.020	&	20	&		&		&		&	39815	&	$5f6d^27p$	&		&	1.050	&		\\
5	&	30372	&		&	1.155	&	30245	&	$5f6d7s7p$	&	127	&	1.142	&	0.013	&	21	&		&		&		&	40601	&	$5f6d^27p$	&		&	1.096	&		\\
6	&	30930	&	$5f6d7s7p$	&	1.2	&	30980	&	$5f6d7s7p$	&	-50	&	1.243	&	-0.043	&	22	&		&		&		&	40834	&	$5f6d^27p$	&		&	1.143	&		\\
7	&	31210	&	$6d^4$	&		&	31186	&	$5f6d7s7p$	&	24	&	1.171	&		&	23	&	41214	&	$5f6d^27p$	&	1.1	&	41303	&	$5f6d^27p$	&	-89	&	1.076	&	0.024	\\
8	&	31716	&		&	1.21	&	31586	&	$6d^4$	&	130	&	1.058	&	0.152	&	24	&		&		&		&	41797	&	$5f6d^27p$	&		&	1.084	&		\\
9	&	33068	&		&	1.08	&	33089	&	$5f6d7s7p$	&	-21	&	1.079	&	0.002	&	25	&		&		&		&	42298	&	$5f6d^27p$	&		&	1.074	&		\\
10	&	33603	&		&	1.205	&	33373	&	$5f6d7s7p$	&	230	&	1.187	&	0.018	&	26	&	42841	&		&	1.17	&	42650	&	$5f6d^27p$	&	191	&	1.190	&	-0.020	\\
11	&	34407	&		&	1.19	&	34312	&	$5f6d7s7p$	&	95	&	1.222	&	-0.032	&	27	&		&		&		&	42798	&	$5f6d^27p$	&		&	0.985	&		\\
12	&	34943	&		&	1.23	&	34872	&	$5f6d7s7p$	&	71	&	1.285	&	-0.055	&	28	&		&		&		&	43265	&	$5f6d^27p$	&		&	0.922	&		\\
13	&	35082	&		&		&	35285	&	$5f6d7s7p$	&	-203	&	1.251	&		&	29	&	43937	&		&	1.155	&	43744	&	$5f6d7s7p$	&	193	&	1.129	&	0.026	\\
14	&	35800	&	$5f6d7s7p$	&	1.09	&	35961	&	$5f6d7s7p$	&	-161	&	1.025	&	0.066	&	30	&		&		&		&	44059	&	$5f^26d7s$	&		&	1.051	&		\\
15	&	36749	&		&	1.13	&	36957	&	$5f6d^27p$	&	-208	&	1.046	&	0.084	&	31	&		&		&		&	44428	&	$5f^26d7s$	&		&	1.139	&		\\
16	&	37332	&	$5f6d^27p$	&	1.02	&	37315	&	$5f6d^27p$	&	17	&	1.061	&	-0.041	&	32	&	44438	&		&	1.125	&	44607	&	$5f6d^27p$	&	-169	&	1.193	&	-0.068	\\

\end{tabular}
\end{ruledtabular}
\end{table*}

\begin{table*}
\caption{Comparison of energies, g factors, and dominant configurations of CI-MBPT with nine optimized parameters (Table II) with Ref.\cite{Thorium}. The energies of states for each J were shifted to remove the average systematic shift. In case of J=7, the alignment was done with the lowest J=6 energy and the J=6 optimized set of parameters was used to predict levels for J=7 without further optimization for these states.  This is done to demonstrate predictive power of the method. Energies are given in cm$^{-1}$.  \label{TableV}}
\begin{ruledtabular}
\begin{tabular}{rccccccccccc}
Energy &   Conf.        &  $g$-factor &  Energy   &   Conf.       &  $g$-factor &  Energy   &   Conf.      &  g factor  &   Energy  &  Conf.       & g factor  \\
       &   Ref.\cite{Thorium}      &           &           &  CI-MBPT      &           &           &    Ref.\cite{Thorium}   &            &           & CI-MBPT      &           \\
 \hline
	    &               &     J=1      &           &               &           &           &              &    J=4        &           &              &           \\																						
3865	&	$6d^27s^2$	&	1.48	&	3406	&	$6d^27s^2$	&	1.494	&	4962	&	$6d^27s^2$	&	1.21	&	5008	&	$6d^27s^2$	&	1.22	\\
5563	&	$6d^37s$	&	0.065	&	5595	&	$6d^37s$	&	0.045	&	8111	&	$6d^27s^2$	&	1.065	&	7823	&	$6d^37s$	&	1.319	\\
11601	&	$6d^37s$	&	2.4	&	11548	&	$6d^37s$	&	2.443	&	8800	&	$6d^37s$	&	1.31	&	8780	&	$6d^27s^2$	&	1.046	\\
13962	&	$6d^37s$	&	0.76	&	14164	&	$6d^37s$	&	0.647	&	13297	&	$6d^37s$	&	1	&	13291	&	$6d^37s$	&	1.012	\\
17074	&	$6d^37s$	&	1.28	&	16904	&	$6d^37s$	&	1.355	&	15493	&	$6d^37s$	&	0.905	&	15726	&	$6d^37s$	&	0.894	\\
18574	&	$6d^37s$	&	1.365	&	19146	&	$6d^37s$	&	1.398	&	17960	&	$6d^37s$	&	1.175	&	17506	&	$6d^37s$	&	1.195	\\
21579	&	$6d^4$	&		&	21617	&	$6d^4$	&	1.428	&	19532	&	$6d^37s$	&	1.204	&	20007	&	$6d^37s$	&	1.182	\\
22401	&	$6d^37s$	&	1.185	&	22239	&	$6d^37s$	&	1.169	&	21645	&	$6d^37s$	&	1.09	&	21734	&	$6d^37s$	&	1.074	\\

\hline
	    &		        &	J=2	&		&		&		&		&		&	J=5	&		&		&		\\
0	&	$6d^27s^2$	&	0.735	&	81	&	$6d^27s^2$	&	0.727	&	9805	&	$6d^37s$	&	1.365	&	9055	&	$6d^37s$	&	1.378	\\
3688	&	$6d^27s^2$	&	1.255	&	3578	&	$6d^27s^2$	&	1.289	&	14204	&	$6d^37s$	&	1.15	&	14399	&	$6d^37s$	&	1.157	\\
6362	&	$6d^37s$	&	1.01	&	6185	&	$6d^37s$	&	1.008	&	17166	&	$6d^37s$	&	1.115	&	17318	&	$6d^37s$	&	1.092	\\
7280	&	$6d^27s^2$	&	1.185	&	6876	&	$6d^27s^2$	&	1.155	&	21143	&	$6d^37s$	&	1.03	&	21609	&	$6d^37s$	&	1.008	\\
11803	&	$6d^37s$	&	1.7	&	11816	&	$6d^37s$	&	1.739	&	23277	&	$5f6d7s7p$	&	1.01	&	23471	&	$5f6d7s7p$	&	0.996	\\
13848	&	$6d^37s$	&	0.945	&	13878	&	$6d^37s$	&	0.925	&	26380	&	$5f6d7s7p$	&	1.025	&	26256	&	$5f6d7s7p$	&	0.99	\\
15863	&	$6d^37s$	&	1.07	&	15788	&	$6d^37s$	&	1.031	&	27191	&	$5f6d7s7p$	&	1.12	&	27054	&	$5f6d7s7p$	&	1.115	\\
18549	&	$6d^37s$	&	1.21	&	18613	&	$6d^37s$	&	1.393	&	27592	&	$5f6d7s7p$	&	1.085	&	27513	&	$5f6d7s7p$	&	1.064	\\

\hline
	&		&	J=3	&		&		&		&		&		&	J=7	&	(opt.J=6) 	&		&		\\
2869	&	$6d^27s^2$	&	1.085	&	2515	&	$6d^27s^2$	&	1.084	&	30727	&	$5f6d7s7p$	&	1.21	&	30007	&	$5f6d7s7p$	&	1.215	\\
7502	&	$6d^37s$	&	1.25	&	7189	&	$6d^37s$	&	1.248	&	32994	&		&	1.235	&	32094	&	$5f6d7s7p$	&	1.236	\\
12848	&	$6d^37s$	&	1.39	&	12659	&	$6d^37s$	&	1.632	&	36468	&		&		&	36172	&	$5f6d7s7p$	&	1.226	\\
13088	&	$6d^37s$	&	1.05	&	13517	&	$6d^37s$	&	0.81	&	36618	&		&	1.185	&	36505	&	$5f6d7s7p$	&	1.205	\\
15970	&	$6d^37s$	&	1.205	&	16032	&	$6d^37s$	&	1.2	&	38703	&	$5f6d7s7p$	&	1.125	&	38169	&	$5f6d^27p$	&	1.19	\\
17398	&	$6d^37s$	&	1.195	&	17433	&	$6d^37s$	&	1.19	&	39600	&	$5f6d7s7p$	&	1.11	&	39328	&	$5f6d^27p$	&	1.161	\\
18431	&	$5f7s^27p$	&		&	18531	&	$5f7s^27p$	&	0.846	&	40281	&		&	1.17	&	39419	&	$5f6d^27p$	&	1.116	\\
19713	&	$6d^37s$	&	1.11	&	20212	&	$6d^37s$	&	1.14	&		&		&		&	40734	&	$5f6d^27p$	&	1.119	\\

\end{tabular}
\end{ruledtabular}
\end{table*}
\subsection{CI-MBPT energies for J=6 even states}
Because fitting J=6 even states resulted in the best agreement with experiment, we have studied this case in more detail (Table~\ref{Table1}). One particular expectation is that if the energies are quite close to the experimental values, then other atomic observables such as g factors will also have good accuracy. This expectation is based on the fact that correct intervals between interacting configurations lead to relatively accurate mixing coefficients of these configurations. The mixing coefficients also strongly influence the matrix elements and other properties. Indeed for many states we find that g factors are in very close agreement with experiment. Apart from being the theory accuracy test parameter, g factors provide an aid in assignment of theoretical levels.
Remarkably, the first level has only 0.2\% deviation and many other agree at about 0.5\% level. Because of such good agreement for energies and g factors, there is some confidence that the levels higher than used in the fitting procedure should have a good accuracy. While unfortunately many levels are missing, the 17th has a good agreement, while some levels like 23rd, 26th can be tentatively assigned to the experimental values by their small deviation and matching g factor. Of course missing levels introduce relatively large uncertainty in the identification, and further work would be needed to experimentally observe them.

\subsection{CI-MBPT energies for J=1-5 and J=7 even states}
The results of CI-MBPT calculations for J=1-5 and J=7 even states using fitted parameters from Table \ref{TableII} are presented in Table~\ref{TableV}. The energies of the experimental and theoretical states are aligned for each J with the systematic average shift removed. The reason for doing this is that the optimization was done separately for each J and the interaction between states of the same symmetry, J, is what affects the mixing coefficients and hence correctness of wavefunctions. Thus relative distances between different J are of less importance than distances between states of the same J. In principles, it is possible to adjust the nine parameters to improve agreement between experiment and theory for the distances between different J states, but then the accuracy within one specific J would be compromised. Still in one case J=7 we align states for J=6, so we can see the prediction of theory when the 9 parameters were optimized for J=6 states, while it is applied to J=7 states. As we can see, the theory still reproduces energies and g factors quite accurately. Furthermore, the theory in similar way can be applied to predict J=8 and J=9 states, most of which are not yet discovered experimentally. For example, the first J=8 state has energy 34559 cm$^{-1}$ and g factor 1.245, which agree with theoretical prediction, 33706 cm$^{-1}$ and 1.2499, while other experimental levels which theory predicts (40623, 41392, etc.) with g factors 1.1958, 1.1990, etc. are not given in \cite{Thorium}. Theory also predicts energies unobserved J=9 states: 43117, 47995 cm$^{-1}$. Thus apart from making identifications with existing experimental data, theory can also predict new levels. The accuracy depends on how similar those states are to the states which were used in fitting. This similarity includes energy value, J, and configurations.

\section{Discussion and Conclusion}

 We found that CI-MBPT energies and g factors agree not only with experiment but with a precision {\it ab initio} CI-All-order theory \cite{ThSafronova}. One important theoretical question is to understand the importance of configuration mixing effects: essentially, would it be better to have {\it ab initio} theory which has some relatively large deviations from experiment or would it be the accuracy improved for some class of problems with the aid of fitting parameters? While various properties are important, the identification of levels is a crucial starting step of theoretical applications. By reducing theoretical error for energies with fitting, the identification becomes much easier. Also we find that g factors are in good agreement too and they are valuable aid in the identification.

 One significant drawback of fitting with many parameters is that the calculations have to be repeated many times to follow minimization algorithm. In case of 9 parameters, the parametric space is huge, so it is just impossible to implement systematic search with a uniform grid. It is also difficult if not impossible to find an absolute minimum and show that it is indeed the lowest minimum among multiple possible local minima. As we already mentioned before, the calculations have to be repeated 100 times to find some local minimum using the algorithm discussed here. This leads to the need to reduce configurational space and hence precision of treatment of valence-valence interactions. On the other hand, once the optimized parameters are found, the other atomic properties can be calculated very fast.

 In the future, work needs to be done to find most efficient algorithms for optimization of parameters. Also the mathematical properties of the energy dependence on parameters needs to be better understood. One simple fact is that the parameters that take into account single valence energies, 1-7, shift energies linearly if the interaction mixing is weak. In this case, the minimum can be found easily just from gradients. However, when two levels approach each other and there is strong interaction between leading configurations of these states, then mixing can become quite large and the dependence of energies of states will become non-linear. The configuration mixing will also strongly affect g factors, so in many cases of two strongly interacting states, the sum of g factors might be preserved while each individual g factor can be quite different from its true value. Other properties, such as matrix elements, are also strongly affected by configuration mixing. In this case some correlation might exist between the accuracy of g factors and matrix elements.

In conclusion, the CI+MBPT method with 9 adjustable parameters was used to calculate energy levels and g factors of Th I. A close agreement was found for energy levels which allowed to match theoretical and experimental levels. Excellent agreement was also observed for g factors for many states presented in the tables.
Finally, it is expected that in the future the CI-MBPT method with multiple adjustable parameters can be further developed to treat most complex atoms, including lantanides and actinides. Considering limitations of {\it ab initio} approaches and high sensitivity of the atomic properties to small corrections, it seems that at some level parametric approach might be the only solution to the problem of obtaining correct actinide properties.

\section{Acknowledgement}
The work of I. Savukov has been performed under the auspices of the U.S.$\sim$DOE by LANL
under contract No.$\sim$DE-AC52-06NA25396. The author is grateful to Dr. Dzuba for making his CI+MBPT code available for this work.

\end{document}